\documentclass[aps,prl,groupedaddress,onecolumn,10pt, superscriptaddress, notitlepage, longbibliography]{revtex4-1}

\usepackage{lipsum}
\usepackage{amsmath}
\usepackage{amssymb}
\usepackage{amsthm}
\usepackage{bbm}
\usepackage{graphicx}
\usepackage{color}
\usepackage{upgreek}
\usepackage[linkcolor = blue, citecolor = blue, urlcolor = blue, colorlinks = true]{hyperref}
\usepackage{amssymb}
\usepackage{bm}
\usepackage[capitalise]{cleveref}
\usepackage{textcomp}
\usepackage{hyperref}
\usepackage[usenames,dvipsnames]{xcolor}
\usepackage[normalem]{ulem}
\usepackage{wasysym}

\newcommand\diff{\mathrm{d}}
\renewcommand{\vec}[1]{\mathbf{#1}}
\renewcommand{\imath}[0]{\mathsf{i}}

\hypersetup{colorlinks=true, linkcolor=blue!50!black, urlcolor=blue!50!black, citecolor=blue!50!black}

\definecolor{ABpurple}{RGB}{128, 0, 128}
\definecolor{ABred}{RGB}{255, 0, 0}
\definecolor{ABgreen}{RGB}{0, 255, 0}
\definecolor{ABbrown}{RGB}{128, 64, 0}
\definecolor{ABblue}{RGB}{0, 0, 255}

\newcommand{\inlinemaketitle}{{\let\newpage\relax\maketitle}}
\begin{document}

\title{Particle motion nearby rough surfaces}

\author{Christina Kurzthaler}
\affiliation{Department of Mechanical and Aerospace Engineering, Princeton University, New Jersey 08544, USA}
\author{Lailai Zhu}
\affiliation{Department of Mechanical and Aerospace Engineering, Princeton University, New Jersey 08544, USA}
\affiliation{Department of Mechanical Engineering, National University of Singapore, 117575, Singapore}
\affiliation{Linn{\'e} Flow Centre and Swedish e-Science Research Centre (SeRC), KTH Mechanics, SE-10044, Stockholm, Sweden}
\author{Amir A. Pahlavan}
\affiliation{Department of Mechanical and Aerospace Engineering, Princeton University, New Jersey 08544, USA}
\author{Howard A. Stone}
\email{hastone@princeton.edu}
\affiliation{Department of Mechanical and Aerospace Engineering, Princeton University, New Jersey 08544, USA}

\begin{abstract}
We study the hydrodynamic coupling between particles and solid, rough boundaries characterized by random surface textures. Using the Lorentz reciprocal theorem, we derive analytical expressions for the grand mobility tensor of a spherical particle and find that roughness-induced velocities vary nonmonotonically with the characteristic wavelength of the surface. In contrast to sedimentation near a planar wall, our theory predicts continuous particle translation transverse and perpendicular to the applied force.
Most prominently, this motion manifests itself in a variance of particle displacements that grows quadratically in time along the direction of the force. This increase is rationalized by surface roughness generating particle sedimentation closer to or farther from the surface, which entails a significant variability of settling velocities.
\end{abstract}

\maketitle
Particle sedimentation in low-Reynolds-number flows represents a fundamental problem in physics and fluid dynamics and has been studied over decades
due to its relevance for natural and technological applications. These range from the separation of multicomponent systems on the microscale~\cite{Wang:2002,Sajeesh:2014} to macroscopic geological phenomena~\cite{Garcia:2008}. Confining geometries omnipresent in  natural environments and microfluidic devices, however, significantly alter the sedimentation process due to long-range hydrodynamic interactions~\cite{Segre:1997,Tee:2002,Ladd:2002,Heitkam:2013}. Unlike planar surfaces, these boundaries display a large variety of structured and rough topographies with random heterogeneities, which modify the surrounding flow fields~\cite{Bonaccurso:2003, Kunert:2010,Kamrin:2010,Lee:2014,Assoudi:2018_1} and hydrodynamically impact nearby particle motion~\cite{Hosseini:2010,Assoudi:2018}. Understanding the physical mechanisms underlying the interactions of the particles with these boundaries lays the foundation for the design of novel particle separation methods~\cite{Huang:2004,Davis:2006,McGrath:2014} and tools for the noninvasive measurement of surface properties reminiscent of microrheology~\cite{Squires:2010,Zia:2018}.

A rigid spherical particle that sediments nearby a vertical planar wall does not translate perpendicular to the wall and thus keeps a constant distance to it~\cite{Oneill:1967,Goldman:1967}. This behavior, however, changes drastically for spheroidal particles and slender rods, whose anisotropic shape generates intricate tumbling behavior~\cite{Russel:1977,Mitchell:2015}, and in the presence of fluid inertia~\cite{Vasseur:1977,Becker:1996} or near elastic boundaries~\cite{Rallabandi:2018}, which can both induce a lift force and generate particle migration away from the surface. The impact of complex surface shapes has only been studied for particles suspended in shear flow near periodic surfaces that can induce motion across streamlines~\cite{Belyaev:2017,Assoudi:2018, Charru:2007}. Despite its ubiquity, however, the hydrodynamic coupling between sedimenting particles and  rough surfaces with random features remains an open question at the interface of low-Reynolds-number flows and statistical physics.

Here, we study hydrodynamic interactions between a sphere and a textured surface and provide an analytic expression for the particle mobility. We employ our theory to elucidate the motion of a sedimenting sphere near a rough surface with random features and find that the velocities depend nonmonotonically on the characteristic wavelength of the surface. In contrast to sedimentation near a planar wall, the  particle translates perpendicular and transverse to the applied force. Our study reveals that these hydrodynamic interactions generate a variability in settling velocities and thereby induce a variance of the particle displacements along the force direction that displays a quadratic increase at long times. We further relate the particle mobility to the spatially dependent diffusivity of a Brownian particle.

\begin{figure}[htp]
\centering
\includegraphics[width=0.6\linewidth]{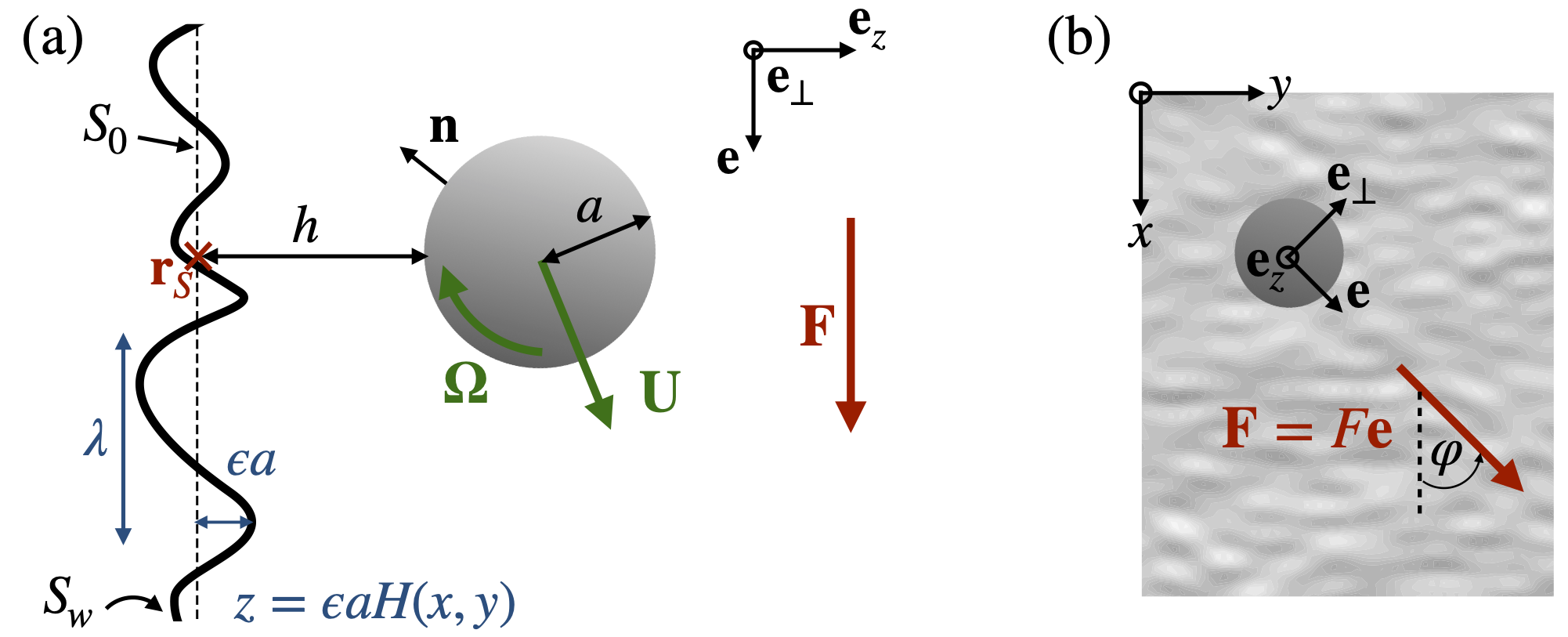}
\caption{(a) Side view of a sphere with radius $a$ near a vertical rough wall, $S_w$, characterized by the shape function $H(x,y)$, a wavelength $\lambda$, and amplitude $\epsilon a$. In response to an external force $\vec{F}=F\vec{e}$ (parallel to $S_0$), the sphere translates and rotates with velocities $\vec{U}$ and $\vec{\Omega}$, respectively. The sphere is located at $\vec{r}_S$ and at a distance $h$ relative to $S_0$. (b) Top view of the sphere and the rough surface (grayscale indicates a height map), where $x,y$ are the coordinates spanning $S_0$. Here,  $\vec{F}$ is oriented at an angle $\varphi$ relative to the surface structure and $\vec{e}$, $\vec{e}_\perp$, and $\vec{e}_z$ denote the unit vectors along, transverse, and perpendicular to~$\vec{F}$.
\label{fig:sketch}}
\end{figure}

\emph{Model.--} We consider the motion of a spherical particle with radius $a$ in an incompressible flow of a viscous fluid nearby a rough surface, $S_w$, in three dimensions ($3$D) [Fig.~\ref{fig:sketch}].
The quasi-steady fluid velocity $\mathbf{u}(\mathbf{r})$ and pressure fields $p(\mathbf{r})$ are described by the continuity and Stokes equations, $\nabla\cdot\vec{u} = 0$ and $\nabla\cdot\boldsymbol{\sigma}=\vec{0}$,
with stress field $\boldsymbol{\sigma} = -p\mathbb{I} + \mu \left(\nabla \vec{u}+\nabla \vec{u}^T\right)$ and viscosity $\mu$.
In the co-moving frame of reference that is attached to the sphere the no-slip boundary conditions (BCs) are:
$\vec{u}= \boldsymbol{\Omega} \wedge \vec{r}$ on $S_p$ and $\vec{u}=-\vec{U}$ on $S_w$ and $S_\infty$, which denotes the bounding surface at infinity. The instantaneous translational and rotational velocities of the sphere, $\vec{U}=\vec{U}(\vec{r}_S,h)$ and $\vec{\Omega}=\vec{\Omega}(\vec{r}_S,h)$, depend on its distance from the wall, $h=h(t)$, 
its position, $\vec{r}_S=\vec{r}_S(t)$, relative to the underlying surface, which determines the particle
velocities locally, and time~$t$.

We describe the small height fluctuations of the textured surface by $z=\epsilon a H(x,y)$, where $H(x,y)$ denotes the shape function  and  $\epsilon a$ the surface amplitude with dimensionless parameter $\epsilon\ll 1$. Assuming that the surface amplitude is smaller than the particle-wall distance,  $\epsilon a\ll h$, we expand the velocity field in $\epsilon$, $\mathbf{u} = \mathbf{u}^{(0)}+\epsilon\mathbf{u}^{(1)} +\mathcal{O}(\epsilon^2)$.
Equivalently, we express the translational and rotational velocities of the sphere by
$\vec{U}=\vec{U}^{(0)}+\epsilon\vec{U}^{(1)}+\mathcal{O}(\epsilon^2)$ and
$\vec{\Omega}=\vec{\Omega}^{(0)}+\epsilon\vec{\Omega}^{(1)}+\mathcal{O}(\epsilon^2)$.
Using the method of domain perturbation, we express the no-slip BC at the rough surface $S_w$ in terms of a Taylor expansion about $z=0$ (surface $S_0$). The expansion enables us to consider the zeroth- and first-order problems separately with BCs, $\vec{u}^{(0)} = -\vec{U}^{(0)}$, and
\begin{align}
\vec{u}^{(1)} &= -\vec{U}^{(1)}-aH(x,y)\frac{\partial \mathbf{u}^{(0)}}{\partial z }\Bigr|_{z=0} \text{ on } S_0, \label{eq:bc_1}
\end{align}
which contains a roughness-induced slip velocity~\cite{Bonaccurso:2003, Kunert:2010,Kamrin:2010} dictated by the zeroth-order flow and the surface shape.

Then the zeroth-order problem with flow field $\mathbf{u}^{(0)}$ corresponds to 
a sphere moving near a planar wall, which has been elaborated analytically in terms of a bispherical representation~\cite{Jeffery:1915,Brenner:1961,Dean:1963,Oneill:1964}.
The first-order correction to the fluid flow, $\mathbf{u}^{(1)}$, encodes details of the rough surface. It obeys the Stokes and continuity equations 
with BCs: $\vec{u}^{(1)}=\vec{\Omega}^{(1)} \wedge \vec{r}$ on $S_p$, $\vec{u}^{(1)}=-\vec{U}^{(1)}$ on $S_\infty$, and  Eq.~\eqref{eq:bc_1}.
In creeping flow, the total force and torque exerted on the particle are zero. Here, the applied force, $\vec{F}$, and torque, $\vec{L}$,
are balanced by the hydrodynamic force and torque of the zeroth-order problem, $\vec{F}_H^{(0)}=\int_{S_p} \vec{n}\cdot\boldsymbol{\sigma}^{(0)} \diff S$ and $\vec{L}_H^{(0)}=\int_{S_p} \vec{r} \wedge \left(\vec{n}\cdot\boldsymbol{\sigma}^{(0)}\right) \diff S$, where the normal vector $\vec{n}$ is directed away from $S_p$.
Consequently, the sphere in the first-order problem is force- and torque-free, which determines its roughness-induced velocities.

\emph{Particle mobility.--} We develop an analytic theory for the mobility of a sphere near a rough wall by employing the Lorentz reciprocal theorem, which relates two Stokes flow problems that share the same geometry but have different boundary conditions~\cite{Leal:2007,Masoud:2019}.
We introduce as the auxiliary problem, $\hat{\vec{u}}$, $\hat{\boldsymbol{\sigma}}$, the flow generated by a moving
sphere near a planar wall. The Lorentz reciprocal theorem relates it to
the first-order problem with velocity field $\vec{u}^{(1)}$ and stresses $\boldsymbol\sigma^{(1)}$ via
\begin{align}
\int_{S_p,S_0,S_\infty} \vec{n}\cdot \boldsymbol{\sigma}^{(1)} \cdot \hat{\vec{u}} \ \diff S &= \int_{S_p,S_0,S_\infty} \vec{n}\cdot \hat{\boldsymbol{\sigma}} \cdot \vec{u}^{(1)} \ \diff S. \label{eq:lorentz}
\end{align}
Since the first-order problem is force- and torque-free, the left-hand side of Eq.~\eqref{eq:lorentz} vanishes. We insert the BCs of the main and auxiliary problems into Eq.~\eqref{eq:lorentz} and note that by the divergence theorem the hydrodynamic force on the sphere in the auxiliary problem obeys $\hat{\vec{F}}_H = -\int_{S_0,S_\infty} \, \vec{n}\cdot \hat{\boldsymbol{\sigma}} \ \diff S$.
Thus, Eq.~\eqref{eq:lorentz} simplifies to
\begin{align}
\hat{\mathbf{F}}_H\cdot\mathbf{U}^{(1)}+\hat{\mathbf{L}}_H\cdot\mathbf{\Omega}^{(1)} =
&\int_{S_0} \, \! aH(x,y) \vec{n}\cdot\hat{\boldsymbol{\sigma}}\cdot\frac{\partial \mathbf{u}^{(0)}}{\partial z }\Bigr|_{z=0} \diff S.
\label{eq:lorentz_simplified}
\end{align}

Due to the linearity of the Stokes equations and the rigid boundaries of the particle and the wall,
the particle velocities
must be coupled linearly to the applied force and torque
via the grand mobility tensor $\vec{M}$, $(\vec{U}, \vec{\Omega})^T=\vec{M}\cdot(\vec{F}, \vec{L})^T$~\cite{Leal:2007}.
Its components encode the coupling between the translational and
rotational velocities with the forces and torques, $\vec{M}_{UF}$ and $\vec{M}_{\Omega L}$, respectively, and between the rotation and force (translation and torque), $\vec{M}_{\Omega F}$ ($\vec{M}_{U L}$).
We further expand the mobility tensor in terms of the small roughness parameter~$\epsilon$,
$\vec{M}= \vec{M}^{(0)}+\epsilon\vec{M}^{(1)}+\mathcal{O}(\epsilon^2)$,
where $\vec{M}^{(0)}$ is the mobility of a sphere near a planar wall and $\vec{M}^{(1)}$  corresponds to the roughness-induced mobility.
Since the velocity field and stresses are linear in the forces and torques,
we arrive at the general form for the grand mobility tensor [see Supplemental Material (SM)~\cite{supplement}]
\begin{align}
\mathbf{M} &= \mathbf{M}^{(0)}-\epsilon \int_{S_0} \, aH(x,y) \vec{K} \ \diff S +\mathcal{O}(\epsilon^2), \label{eq:mobility}
\end{align}
which accounts for the influence of surface topography.
The coupling tensor $\vec{K}$ can be assumed known as it depends on the zeroth-order problem only. The grand mobility tensor represents an exact result up to second order in $\epsilon$ determined by the (arbitrary)
surface shape and the instantaneous position of the sphere, 
$\vec{M}[\vec{r}_S(t),h(t);H]$. The linearity of the Stokes equations and the rigid boundaries entail that $\bf M$ is symmetric and positive definite~\cite{Kim:2013}.

We note that the analytical theory is valid for low-Reynolds-number flows and small surface roughness with Re~$\lesssim\epsilon\lesssim1$. For Re~$\sim\epsilon$ our theory needs to be modified to account for inertial effects 'in the spirit of Saffman' that could generate  migration away from the wall~\cite{Vasseur:1977}.

Subsequently, we investigate the sedimentation of a sphere in response to a force, $\vec{F}$, near a random, rough wall. We use the exact bispherical representation of the fluid flow $\vec{u}^0$ to calculate the coupling tensor $\vec{K}$ and evaluate the velocities, $\vec{U}= \mathbf{M}_{UF}\cdot\vec{F}$ and $\vec{\Omega}= \mathbf{M}_{\Omega F}\cdot\vec{F}$,  via numerical integration.
We validated our analytical solutions for a periodic surface with a boundary integral method that captures the full surface shape (see SM~\cite{supplement}).
\begin{figure*}[htp]
\includegraphics[width = \linewidth]{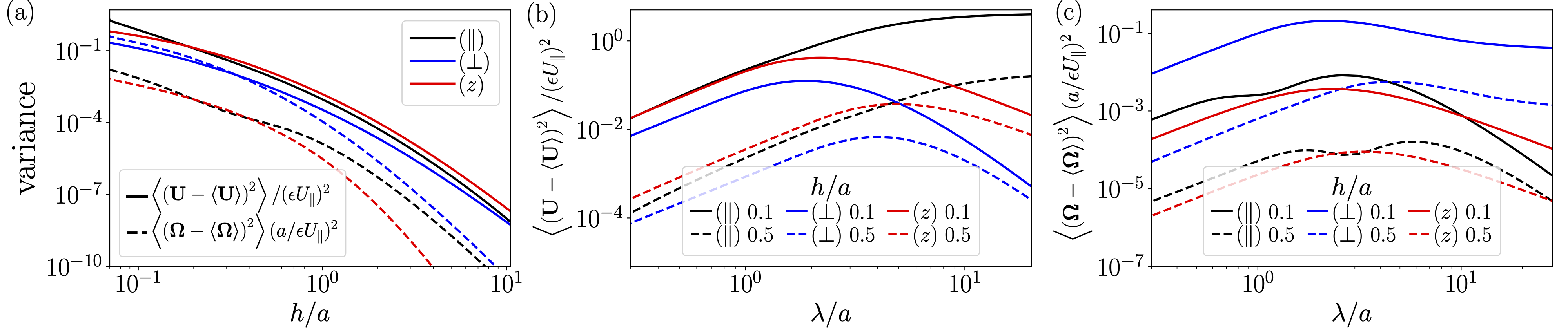}
\caption{Variance of the translational, $\left \langle\! \left(\vec{U}\!-\!\left\langle\vec{U}\right\rangle\right)^2\right\rangle$, and rotational velocities, $\left \langle\! \left(\vec{\Omega}\!-\!\left\langle\vec{\Omega}\right\rangle\right)^2\right\rangle$,
with respect to (a) the distance $h/a$  with $\lambda/a = 2$ and (b), (c) the wavelength $\lambda/a$. (b) and (c) show the variances for different particle-surface distances, $h/a=0.1,0.5$.
Here, $U_\parallel$ denotes the settling velocity of a sphere near a planar wall and we have used $N=50$ in Eq.\eqref{eq:H_rough}. The variances are shown for the velocities along $(\parallel)$, transverse $(\perp)$, and perpendicular $(z)$ to the force direction, $\vec{F}=F\vec{e}_x$.
\label{fig:var}}
\end{figure*}

\emph{Roughness-induced velocities.--}
We describe the random rough wall by a statistical framework~\cite{Savva:2010}.
The surface shape is modeled as a superposition of $N\times N$ Fourier modes with random amplitudes $\alpha_{nm}$, $\beta_{nm}$,
\begin{align}
\begin{split}
H(\vec{r}_0)\!=\!\frac{1}{N}\!\!\sum_{n,m=1}^N\!\!\!\alpha_{nm}\sin\left(\vec{k}^n_{m}\!\cdot\!\vec{r}_0\right)\!+\!\beta_{nm} \cos\left(\vec{k}^n_{m}\!\cdot\!\vec{r}_0\right),\!
\end{split}
\label{eq:H_rough}
\end{align}
where $\mathbf{r}_0\in S_0$, $\vec{k}^n_{m}=(k_n,k_m)^T$ with $k_{j}=2\pi j/(\lambda N)$, and wavelength $\lambda$.
The amplitudes are statistically independent, random normal variables with zero mean and unit variance.
The height fluctuations vanish on average, $\langle H\rangle = 0$, and display a variance, $\langle (\epsilon a H)^2\rangle = (\epsilon a)^2$, where $\langle\cdot\rangle$ is the average over all surface realizations.

The applied force, $\vec{F}=F\vec{e}$, is directed parallel to $S_0$ but can be oriented with respect to the surface structure [Fig.~\ref{fig:sketch}(b)].
Then roughness-induced velocities are decomposed into components parallel, transverse, and perpendicular to it, $\vec{U}^{(1)}= U_\parallel^{(1)}\vec{e}+U_\perp^{(1)}\vec{e}_\perp+U_z^{(1)}\vec{e}_z$ with $\vec{e}_\perp=\vec{e}_z \wedge \vec{e}$ and similarly
$\vec{\Omega}^{(1)}$. They are determined by the local geometry of the underlying, random surface, and, therefore, we are interested in average quantities. Since the velocities depend linearly on the surface shape, the average velocities remain independent of the roughness and reduce to that of a sphere near a planar wall, $\langle \vec{U}\rangle =\vec{U}^{(0)}=U_\parallel \vec{e}$ and $\langle \vec{\Omega}\rangle =\vec{\Omega}^{(0)}=\Omega^{(0)}\vec{e}^\perp$, where $\Omega^{(0)}$ is related to $U_\parallel$.
The settling velocity decreases for decreasing $h/a$ due to the wall-induced extra hydrodynamic drag and is obtained via the balance of gravitational and hydrodynamic forces, $U_\parallel\sim \ln(a/h)^{-1}$ (for $h/a\lesssim1$)~\cite{Oneill:1967}.
The sphere rotates to balance the hydrodynamic torque arising from the presence of the wall~\cite{Goldman:1967}.

The velocity fluctuations due to the surface roughness become apparent in the variance of the translational velocities, which grows quadratically to leading order in~$\epsilon$, $\langle\! (\vec{U}\!-\!\left\langle\vec{U}\right\rangle)^2\rangle = \epsilon^2\langle\! (\vec{U}^{(1)})^2\rangle +\mathcal{O}(\epsilon^3)$
(see SM~\cite{supplement}).
They are determined by the particle-wall distance $h/a$  and wavelength of the surface $\lambda/a$.
The translational velocity fluctuations decay roughly eight orders in magnitude for $0.1\lesssim h/a\lesssim10$  [Fig.~\ref{fig:var}(a)], which indicates that for increasing $h/a$ transport is governed by the average wall contribution. We note that the velocity fluctuations transverse to the force are smaller than perpendicular to it.
The variance of the rotational velocities, $\langle\! (\vec{\Omega}\!-\!\left\langle\vec{\Omega}\right\rangle)^2\rangle = \epsilon^2\langle\! (\vec{\Omega}^{(1)})^2\rangle +\mathcal{O}(\epsilon^3)$, decays even faster and remains most pronounced for rotation around~$\vec{e}_\perp$.

Most prominently, the variances of the perpendicular, $\langle (U^{(1)}_z)^2\rangle$, and transverse velocities, $\langle (U^{(1)}_\perp)^2\rangle$, display a maximum at a characteristic wavelength $\lambda_\text{max}$ [Fig.~\ref{fig:var}(b)].
This nonmonotonic behavior can be rationalized as for $\lambda\lesssim\lambda_\text{max}$ the surface area closest to the particle contains several heterogeneities with steep slopes
that smear out roughness-induced flows and therefore the velocity fluctuations  perpendicular and transverse to the force vanish.
In contrast, for $\lambda\gtrsim\lambda_\text{max}$ the particle experiences the presence of a smooth wall as the surface slope becomes negligible and thus the roughness-induced velocities decrease. We also find that $\lambda_\text{max}$ for $h/a=0.5$ is $\sim 2$ times larger than for $h/a=0.1$, because larger flows are required to  transmit information of the roughness to a particle that is located further away and these are generated in larger cavities.

Differently, the velocity fluctuations along the force, $\langle (U^{(1)}_\parallel)^2\rangle$, saturate for large $\lambda$, where the surface appears almost flat [Fig.~\ref{fig:var}(b)] but is shifted a small amount closer to or farther away from the sphere.
Then the settling velocity of a sphere near a wall that is  $\epsilon a$ closer to or farther away,
provides an upper bound of the variance
$(U_\parallel(h/a\pm \epsilon)-U_\parallel(h/a))^2\sim(\epsilon a/h)^2U_\parallel(h/a)^2$ (for $h/a\lesssim1$).

The rotational velocity fluctuations display nonmonotonic behaviors where $\langle (\Omega_\perp^{(1)})^2\rangle$
saturates towards a plateau at large $\lambda$ [Fig.~\ref{fig:var}(c)], which can be explained by the same physical mechanisms as introduced before. However, $\langle (\Omega_\parallel^{(1)})^2\rangle$ exhibits a bimodal shape with a local minimum at intermediate $\lambda$, where the flow reflected from nearby surface bumps prevents the sphere from rotating.

\begin{figure}[htp]
\includegraphics[width = 0.75\linewidth]{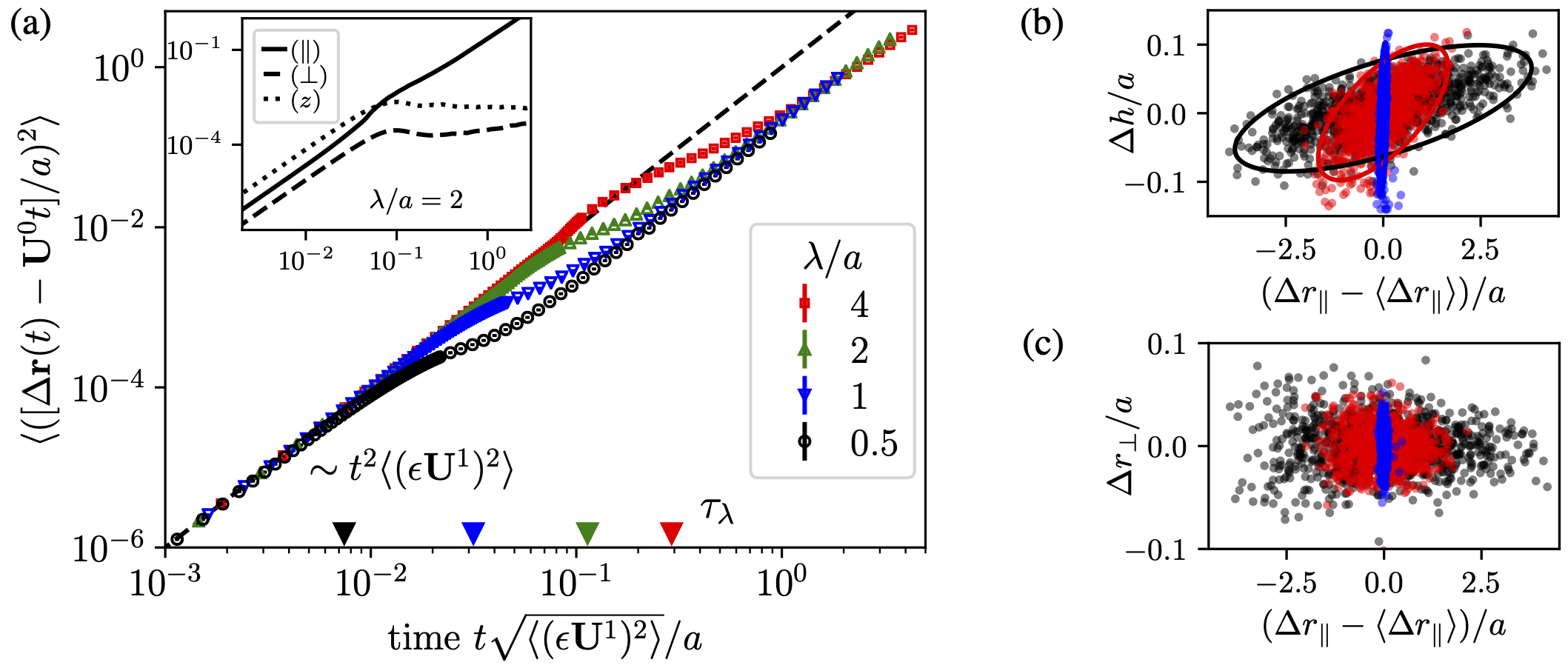}
\caption{(a) Variance of the displacements, $\langle [\Delta \vec{r}(t)-\vec{U}^{(0)}t]^2\rangle$, for different wavelengths $\lambda/a$, $h(0)/a=0.2$, $\epsilon = 0.1$, and $\vec{F}=F\vec{e}=F(\cos\pi/4,\sin\pi/4,0)^T$.
 Over the simulation horizon the sphere has translated $\sim 60\times a$ along the force. Results are obtained by averaging over $10^3$ trajectories along different surface realizations (i.e., Eq.~\eqref{eq:H_rough} with random $\alpha_{nm},\beta_{nm}$ ($N=20$)). Error bars remain smaller than the symbol size. (\emph{Inset}) Variance parallel $(\parallel)$, transverse $(\perp)$, and perpendicular $(z)$ to the force for $\lambda/a=2$.
(b), (c) Distributions of the displacements at subsequent times $t_i$ (blue, red, and black correspond to the smallest, intermediate, and largest $t_i$) for $\lambda/a=2$.
(b) $\Delta h$ and (c) $\Delta r_\perp$ relative to $\Delta r_\parallel-\langle\Delta r_\parallel\rangle$ with $\langle\Delta r_\parallel\rangle=U_\parallel t_i$. The ellipses enclose $95\%$ of the displacements.
\label{fig:msd}}
\end{figure}
\emph{Particle sedimentation near a rough wall.--}
We calculate particle trajectories by numerically integrating the equation of motion, $\dot{\vec{r}}(t) = \vec{U}=\vec{U}^{(0)}+\epsilon\vec{U}^{(1)}$,
where $\vec{r}=(\vec{r}_S, h)^T = r_\parallel \vec{e}+ r_\perp \vec{e}_\perp+ h\vec{e}_z$.
We note that lubrication forces prevent the particle from touching $S_0$.
The system displays two characteristic times scales:  the time the sphere requires for (1) $\tau_\lambda = \lambda/U_\parallel$ passing by a surface bump of wavelength $\lambda$ along the force direction and (2) $\tau = a/\langle (\epsilon \vec{U}^{(1)})^2\rangle^{1/2}$ to move its radius due to hydrodynamic coupling with the surface texture.
We find that by averaging over many surface realizations the particle displacement near a rough wall reduces to that near a planar wall, $\langle \Delta \vec{r}(t)\rangle = \vec{U}^{(0)}t$ with $\Delta \vec{r}(t)=\vec{r}(t)-\vec{r}(0)$.

The impact of surface roughness becomes apparent in the variance of the displacements [Fig.~\ref{fig:msd}(a)], which increases quadratically at short times $t\lesssim\tau_\lambda$ as the sphere sediments by the first surface heterogeneity, $\langle [\Delta \vec{r}(t)-\vec{U}^{(0)}t]^2\rangle \simeq \langle (\epsilon\vec{U}^{(1)})^2\rangle t^2$.
This regime is followed by a superdiffusive regime at $t\sim\tau_\lambda$, which reflects that while sedimenting near the textured surface the particles move around and up/down the underlying obstacles.
By inspecting the individual contributions  [Fig.~\ref{fig:msd}(a) (\emph{inset})], we find that the mean-square displacement (MSD) for the perpendicular motion, $\langle[\Delta h(t)]^2\rangle$, saturates towards a plateau at long times, $t\gtrsim \tau_\lambda$. This result indicates that transport towards and away from the wall is confined within a fluid layer near the surface, which is determined by $h(0)$ and  $\lambda$. This response means that the fundamental requirement of reversibility in Stokes flow is preserved.
The MSDs in the transverse direction, $\langle[\Delta r_\perp(t)]^2\rangle$, exhibit a plateau at intermediate times, which indicates that subsequent obstacles bring the particle back to its initial position, so that over the simulation horizon the particle has merely displaced $\sim0.02a$ away from it. This could be due to the periodic nature of the surface structure, which causes, e.g., trajectories around obstacles that display fore-and-aft symmetries.

At long times, $t\gtrsim\tau_\lambda$, the variance exhibits again a ballistic increase [Fig.~\ref{fig:msd}(a)].
The direct contribution due to transverse and perpendicular motions is negligible, yet, due to that motion the settling velocity changes and contributes to the variance of the displacements. Although all particles have started at the same distance $h(0)$, due to the underlying surface structure some particles sediment on average at a distance $h^\star\lesssim h(0)$ closer to the surface and others further away $h^\star\gtrsim h(0)$, so that they display slower, $\vec{U}^{(0)}(h^\star)\lesssim \vec{U}^0[h(0)]$, or faster settling velocities, $\vec{U}^{(0)}(h^\star)\gtrsim \vec{U}^0[h(0)]$. Thus, particles that are closer to the surface displace less than average displacements
$\Delta r_\parallel\lesssim \langle \Delta r_\parallel\rangle$  while others sediment a longer distance than average, $\Delta r_\parallel \gtrsim \langle \Delta r_\parallel\rangle$ [Fig.~\ref{fig:msd}(b)]. Our simulations show that these effects generate a distribution of particle displacements along the force direction of width $\propto t^2$. Moreover, particles that sediment closer to the surface displace farther along the direction transverse to the force [Fig.~\ref{fig:msd}(c)]. Yet, the transverse displacements remain rather small for the times considered, $|\Delta r_\perp| \lesssim 0.1a$. 

\begin{figure}[htp]
\centering
\includegraphics[width = 0.7\linewidth]{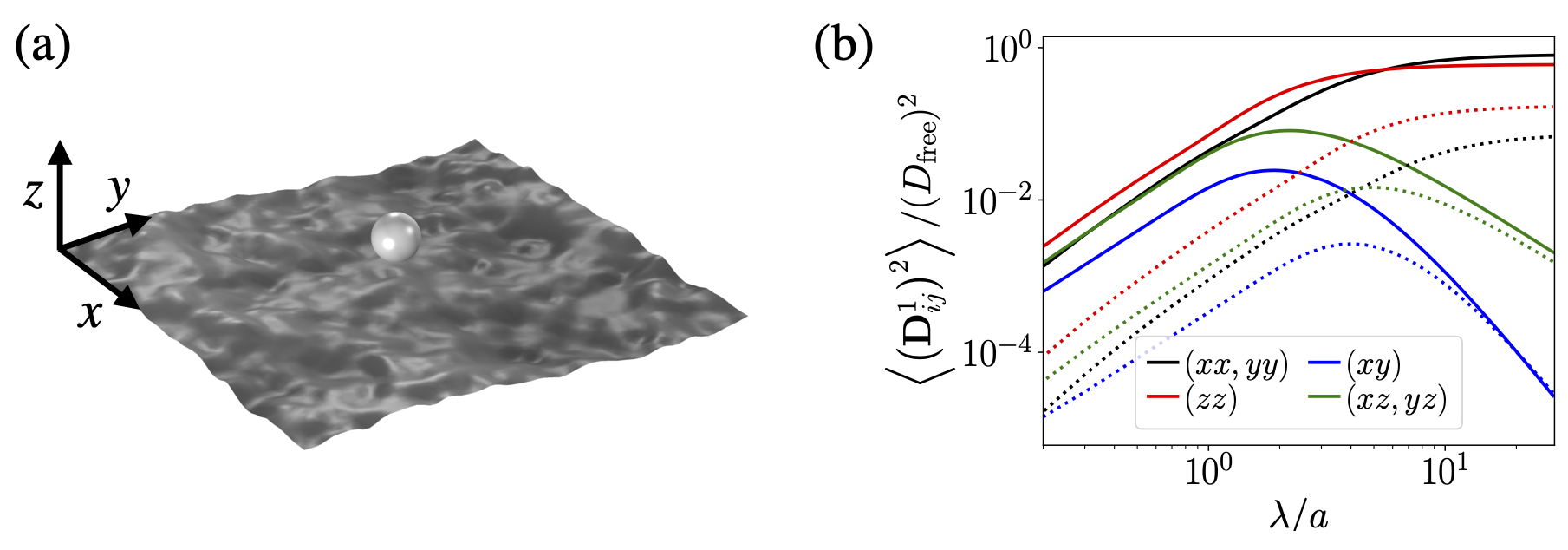}
\caption{(a) Schematic of a sphere diffusing near a rough surface. (b) Variance of the roughness-induced diffusivities,  $\langle(\mathbf{D}^{(1)}_{ij})^2\rangle$, as a function of $\lambda/a$ for $h/a=0.1,0.5$ (solid/dashed lines). The diffusivity in bulk is $D_\text{free}=k_BT/(6\pi\mu a)$.
\label{fig:diff}}
\end{figure}

\emph{Remark on diffusion.--} We note that the hydrodynamic mobility of the sphere is related to the diffusivity of a Brownian particle via the Stokes-Einstein relation, $\mathbf{D}=k_BT\mathbf{M}_{UF}$, where $k_B$ denotes the Boltzmann constant and $T$ temperature.
While a planar wall leads to a well-characterized anisotropic diffusion, where diffusion perpendicular to it is impeded more strongly than parallel motion~\cite{Bian:2016},
a rough surface generates additional, complex spatial dependence of the diffusivities [Fig.~\ref{fig:diff}(a)]. The fluctuations of the roughness-induced diffusivities, $\langle (\mathbf{D}^{(1)}_{ij})^2\rangle=(k_B T)^2\langle [(\mathbf{M}^{(1)}_{UF})_{ij}]^2\rangle$, along the principal axes, parallel [($xx$), ($yy$)] and perpendicular to the surface ($zz$),
increase with $\lambda$ and saturate towards a plateau [Fig.~\ref{fig:diff}(b)].
The rough surface generates correlated motion encoded in the off-diagonal components [($xy$), ($xz$), ($yz$)]
that vary nonmonotonically with~$\lambda$. The underlying physics has been rationalized earlier. For random rough surfaces diffusivities parallel to the surface coincide $\langle (D_{xx}^{(1)})^2\rangle =\langle (D_{yy}^{(1)})^2\rangle$ and thus $\langle(D_{xz}^{(1)})^2\rangle=\langle (D_{yz}^{(1)})^2\rangle$.

\emph{Conclusion.--} We have presented an analytical expression for the mobility of a sphere near a textured surface and studied particle sedimentation near a random, rough wall. Our results show that hydrodynamic interactions between the particle and the surface roughness induce particle translation perpendicular and transverse to the force at velocities that depend nonmonotonically on the wavelength of the surface.
This motion generates a quadratic increase in the variance of the displacements at long times along the force direction,
as particles closer to the surface sediment significantly slower than particles that are farther away.
These results are reminiscent of earlier predictions on force-induced dispersion of Brownian particles in heterogeneous media~\cite{Guerin:2015}, yet they rely on different mechanisms, i.e., hydrodynamic interactions of non-Brownian particles with random, rough surfaces.

Our findings relate statistical transport features of sedimenting non-Brownian particles to the random nature of nearby surface structures that are produced by hydrodynamic coupling. We anticipate that our predictions will allow the noninvasive inference of surface properties by monitoring the transport of particles near rough surfaces, in the same spirit as particle suspensions have been used to quantify the effect of slip heterogeneities in channels~\cite{Vayssade:2014} or diffusivities of Brownian tracers have provided measures of surface slippage~\cite{Lauga:2005, Joly:2006}.
Specifically, measurement of the particle displacements or quantitative extraction of its velocity fluctuations encode information about the characteristic wavelength and amplitude of the surface structure. The roughness-induced mobilities hold for arbitrary surface shapes, which include other random rough surface models or grooved, periodic structures. The latter could serve as input for novel particle separation methods~\cite{Asmolov:2015} that harness the nonmonotonic behavior of particle velocities near textured walls.

Furthermore, our predictions for the particle diffusivities can be employed to explore diffusion near corrugated substrates~\cite{Bian:2016}
and thereby draw a connection to coarse-grained theories~\cite{Reister:2007,Naji:2007}.
Our theory can immediately be extended to derive roughness-induced velocities of microswimmers and thus elucidate the hydrodynamic effect of textured surfaces on active transport~\cite{Takagi:2014,Ostapenko:2018,Frangipane:2019, Makarchuk:2019}.

\begin{acknowledgments}
\paragraph*{Acknowledgments.--} C.K. acknowledges support from the Austrian Science Fund (FWF) via the Erwin Schr{\"o}dinger fellowship (Grant No. J4321-N27). The work was supported by NSF MCB-1853602 (HAS). L.Z. thanks the Swedish Research Council for a VR International Postdoc Grant (2015-06334) and NUS for the startup grant (R-265-000-696-133).
\end{acknowledgments}
\bibliography{literature}

\end{document}


\title{Particle motion nearby rough surfaces: Supplemental information} 

\author{Christina Kurzthaler}
\affiliation{Department of Mechanical and Aerospace Engineering, Princeton University, New Jersey 08544, USA}
\author{Lailai Zhu}
\affiliation{Department of Mechanical and Aerospace Engineering, Princeton University, New Jersey 08544, USA}
\affiliation{Department of Mechanical Engineering, National University of Singapore, 117575, Singapore}
\affiliation{Linn{\'e} Flow Centre and Swedish e-Science Research Centre (SeRC), KTH Mechanics, SE-10044, Stockholm, Sweden}
\author{Amir A. Pahlavan}
\affiliation{Department of Mechanical and Aerospace Engineering, Princeton University, New Jersey 08544, USA}
\author{Howard A. Stone}
\email{hastone@princeton.edu}
\affiliation{Department of Mechanical and Aerospace Engineering, Princeton University, New Jersey 08544, USA}

\maketitle
\setcounter{equation}{0}
\renewcommand\theequation{S.\arabic{equation}}
\renewcommand\thefigure{S.\arabic{figure}}
\section{Particle mobility near a rough wall - derivation of the coupling tensor $\mathbf{K}$}
Due to the linearity of the Stokes equations and the rigid boundaries of the suspended particle and the rough wall, the velocity field generated by the particle motion must be coupled linearly to the applied forces and torques. Consequently, the associated translational and rotational velocities are related to the forces and torques of the problem via the grand mobility tensor $\vec{M}$,
\begin{align}
\begin{pmatrix}\vec{U}\\\vec{\Omega}\end{pmatrix} &= \vec{M}\cdot\begin{pmatrix}\vec{F}\\\vec{L}\end{pmatrix} =\begin{pmatrix}\vec{M}_{UF} & \vec{M}_{UL} \\\vec{M}_{\Omega F} & \vec{M}_{\Omega L}\end{pmatrix}\cdot\begin{pmatrix}\vec{F}\\\vec{L}\end{pmatrix}.
\end{align}
The components of the grand mobility tensor encode the coupling between the translational and rotational velocities with the forces and torques, respectively.
As consequence of the linearity of the Stokes equations, the grand mobility tensor for rigid particles in unbounded configurations, as well as bounded by rigid walls, is always symmetric and positive definite~\cite{Kim:2013}.
We further expand the mobility tensor in terms of the small roughness parameter $\epsilon$,
$\vec{M}= \vec{M}^{(0)}+\epsilon\vec{M}^{(1)}+\mathcal{O}(\epsilon^2)$,
where we have introduced the mobility tensors corresponding to the zeroth- and first-order problems, $\vec{M}^{(0)}$ and $\vec{M}^{(1)}$, respectively. In particular, the mobility tensor for Stokes flow induced by a particle near a smooth plane wall encodes coupling between the force and rotation, and similarly, the torque and translation of the particle, and these couplings are symmetric $\vec{M}^{(0)}_{UL}=\left(\vec{M}^{(0)}_{\Omega F}\right)^T$. Thus, in response to an applied force, in addition to translation, the particle also rotates and vice versa~\cite{Leal:2007}.

Similarly, the velocity field $\vec{u}^{(0)}$ generated by a sphere near a plane wall is linearly related to the applied force and torque via $\vec{u}^{(0)}=\boldsymbol{\mathcal{G}}^{(0)}_F\cdot\vec{F} + \boldsymbol{\mathcal{G}}^{(0)}_L \cdot\vec{L}$, where $\boldsymbol{\mathcal{G}}^{(0)}_F$ and $\boldsymbol{\mathcal{G}}^{(0)}_L$ depend on the distance to the wall, $h$.
Since the normal vector in Eq.~$(5)$ (main text) is for the surface $S_0$, it is directed along the $z-$direction, $\vec{n}=\vec{e}_z$, and the $z-$component of the velocity gradient vanishes by continuity at the wall, i.e. $\vec{e}_z\cdot \partial_z\vec{u}^{(0)}\bigr|_{z=0}=0$, and only off-diagonal elements of the auxiliary stress tensor $\hat{\boldsymbol{\sigma}}$ contribute to the reciprocal relation. Therefore, it can be replaced by the rate of strain tensor
$2\mu\hat{\vec{E}}=-\hat{\boldsymbol{\mathcal{E}}}_F\cdot\hat{\vec{F}}_H -\hat{\boldsymbol{\mathcal{E}}}_L\cdot\hat{\vec{L}}_H$,
where we have defined the third-rank tensors $\hat{\boldsymbol{\mathcal{E}}}_{F,L}$ relating the stresses to the forces and torques.

By linearity we can re-express the reciprocal relation [Eq.~$(5)$ (main text)] and observe that the forces and torques can be eliminated using Eq.~$(6)$ (main text). In particular, we arrive at the general form for the grand mobility tensor of the sphere next to a rough wall
\begin{align}
\mathbf{M} &= \mathbf{M}^{(0)}-\epsilon \int_{S_0} \, aH(x,y) \vec{K} \ \diff S +\mathcal{O}(\epsilon^2),
\end{align}
which represents one of the principal findings of this work. The coupling tensor $\vec{K}$ can be assumed known as it depends on the zeroth-order problem only,
\begin{align}
\vec{K} &= \begin{pmatrix}\left(\vec{n}\cdot\hat{\boldsymbol{\mathcal{E}}}_F\right)^T \cdot \partial_z\boldsymbol{\mathcal{G}}^{(0)}_F &
\left(\vec{n}\cdot\hat{\boldsymbol{\mathcal{E}}}_F\right)^T \cdot \partial_z\boldsymbol{\mathcal{G}}^{(0)}_L \\
\left(\vec{n}\cdot\hat{\boldsymbol{\mathcal{E}}}_L\right)^T \cdot \partial_z\boldsymbol{\mathcal{G}}^{(0)}_F &
\left(\vec{n}\cdot\hat{\boldsymbol{\mathcal{E}}}_L\right)^T \cdot \partial_z\boldsymbol{\mathcal{G}}^{(0)}_L\end{pmatrix}.
\end{align}

\section{General expressions for the roughness-induced velocities}
We consider the motion of a sphere near a rough wall in the presence of an external force, $\vec{F}$.
To evaluate the translational velocities, it is convenient to introduce a local cylindrical coordinate system, $(r,\vartheta,z)$, where $r=\sqrt{(x-x_S)^2+(y-y_S)^2}$ denotes the distance measured from the coordinate of the surface closest to the particle and $\vartheta$ the polar angle. We note that the velocity fields corresponding to the zeroth-order and the auxiliary problems, only depend on the distance $r$ to the sphere and its distance $h$ to the surface, as the problems are translationally invariant with respect to the plane wall. Contrary, the effective slip velocity due to the wall roughness is determined by the underlying surface shape and depends on the instantaneous particle position $\vec{r}_S$. The surface shape measured from the position of the sphere reads
\begin{align}
H(x,y)&=H(x_S+r\cos\vartheta,y_S+r\sin\vartheta)\equiv H(r,\vartheta; \vec{r}_S).
\end{align}

Since the $z$-component of the velocity gradient vanishes by continuity $\vec{e}_z\cdot \partial_z\vec{u}^{(0)}\bigr|_{z=0}=0$, we decompose it in a radial and an angular component via
\begin{align}
\frac{\partial \vec{u}^{(0)}}{\partial z}\Bigr|_{z=0} &= u^{(0)}_{r;z}\vec{e}_r + u^{(0)}_{\vartheta;z}\vec{e}_\vartheta. \label{eq:grad_vel0}
\end{align}

\emph{Translational velocities.} Here, we obtain the roughness-induced velocities by evaluating Eq.~$(5)$ (main text). We note that similarly the full mobility matrix can be calculated, which is not shown here. We introduce the auxiliary problems, which correspond to a sphere that translates along the $x-$, $y-$, and $z-$axis, respectively. We denote the associated stress tensors by $\hat{\boldsymbol{\sigma}}^x$, $\hat{\boldsymbol{\sigma}}^{y}$, and $\hat{\boldsymbol{\sigma}}^{z}$.
Then, the roughness-induced velocities are calculated by evaluating the integrals,
\begin{subequations}
\begin{align}
U_x^{(1)} &= \frac{a}{\hat{F}_H^x}\int_0^\infty\!\!\!\int_0^{2\pi}\, H(r,\vartheta; \vec{r}_S) \left(\hat{\sigma}^x_{zr}u^{(0)}_{r;z} +\hat{\sigma}^x_{z\vartheta}u^{(0)}_{\vartheta;z}\right) r \ \diff\vartheta\diff r, \label{eq:U_parallel}\\
U_y^{(1)} &= \frac{a}{\hat{F}_H^y}\int_0^\infty\!\!\!\int_0^{2\pi}\, H(r,\vartheta; \vec{r}_S)  \left(\hat{\sigma}^y_{zr} u^{(0)}_{r;z}+\hat{\sigma}^y_{z\vartheta} u^{(0)}_{\vartheta;z}\right) r \ \diff\vartheta\diff r, \label{eq:U_y}\\
U_z^{(1)} &= \frac{ a}{\hat{F}_H^z}\int_0^\infty\!\!\!\int_0^{2\pi}\, H(r,\vartheta; \vec{r}_S)  \hat{\sigma}^{z}_{zr} u^{(0)}_{r;z} \  r \ \diff\vartheta\diff r . \label{eq:U_perp}
\end{align}
\end{subequations}
The stress components reduce to $\hat{\sigma}^x_{zr}=\mu\partial_z \hat{u}^x_r$, $\hat{\sigma}^x_{zr}=\mu\partial_z \hat{u}^x_\vartheta$, and $\hat{\sigma}^y_{zr}=\mu\partial_z \hat{u}^y_r$, $\hat{\sigma}^y_{zr}=\mu\partial_z \hat{u}^y_\vartheta$ at $S_0$, respectively.
We have also used that the motion of a sphere away from the wall is axisymmetric and therefore the angular stress component of the auxiliary problem vanishes, $\hat{\sigma}^{z}_{z\vartheta}=0$. The remaining component simplifies to $\hat{\sigma}^{z}_{zr}=\mu\partial_z \hat{u}^z_r$ at $S_0$. 

\emph{Rotational velocities.} Similarly, we obtain the roughness-induced rotational velocities:
\begin{subequations}
\begin{align}
\Omega_x^{(1)} &= \frac{ a}{\hat{L}_H^x}\int_0^\infty\!\!\!\int_0^{2\pi}\! H(r,\vartheta; \vec{r}_S)\left(\hat{\sigma}^{Rx}_{zr} u^{(0)}_{r;z}+\hat{\sigma}^{Rx}_{z\vartheta} u^{(0)}_{\vartheta;z}\right) r \ \diff\vartheta\diff r \label{eq:omega_x}\\
\Omega_y^{(1)} &= \frac{a}{\hat{L}_H^y}\int_0^\infty\!\!\!\int_0^{2\pi}\! H(r,\vartheta; \vec{r}_S)\left(\hat{\sigma}^{Ry}_{zr} u^{(0)}_{r;z}+\hat{\sigma}^{Ry}_{z\vartheta} u^{(0)}_{\vartheta;z}\right) r \ \diff\vartheta\diff r\\
\Omega_z^{(1)} &= \frac{a}{\hat{L}_H^z}\int_0^\infty\!\!\!\int_0^{2\pi}\!  H(r,\vartheta; \vec{r}_S)\hat{\sigma}_{z\vartheta}^{Rz} u_{\vartheta;z}^{(0)} r \ \diff\vartheta\diff r \label{eq:omega_z}
\end{align}
\end{subequations}
Here, the stresses that correspond to the auxiliary problem of a sphere rotating around the $x-$, $y-$, and $z-$axes are denoted by
$\boldsymbol{\sigma}^{Rx}$, $\boldsymbol{\sigma}^{Ry}$, and $\boldsymbol{\sigma}^{Rz}$, respectively.

\section{Zeroth-order problem:  motion of a sphere near a planar wall}
We consider the motion of a sphere driven by an external force, $\vec{F}=F\vec{e}_x$, which is directed parallel to the planar wall $S_0$ along the $x-$direction (we also comment on the general case of a force along an arbitrary direction parallel to $S_0$, $\vec{F}= F\vec{e}$). In addition to translation at velocity $U_\parallel\vec{e}_x$ along the direction of the force, the sphere rotates at a rotational velocity $\vec{\Omega}=\Omega^{(0)} \vec{e}_y$ to balance the hydrodynamic torque generated by the presence of the wall~\cite{Goldman:1967}. Thus, the flow field produced by the motion of the sphere can be decomposed as
\begin{align}
\vec{u}^{(0)} &= \vec{u}^{(0)}_T+\vec{u}^{(0)}_R,
\end{align}
where $\vec{u}^{(0)}_T$ and $\vec{u}^{(0)}_R$ denote the flow fields produced by translation and rotation of the sphere, respectively. We can calculate the zeroth-order flow field by considering the translational and rotational contributions separately. These represent well known problems that have been elaborated earlier in terms of general solutions in bispherical coordinates (Refs. ~\cite{ONeill:1964,Brenner:1961} for translation and Refs.~\cite{Jeffery:1915,Dean:1963} for rotation), in the lubrication limit (Refs.~\cite{Oneill:1967,Goldman:1967,Cox:1967}), where the distance between the sphere and the wall is small, $h/a\lesssim1$, and in the far-field regime, $h/a\gtrsim1$~\cite{Blake:1974}.

The balance of forces and torques determines the translational and rotational velocities, $U_\parallel$ and $\Omega^{(0)}$, of the driven sphere. In particular, the total hydrodynamic force and torque on the sphere, $\vec{F}_H^{(0)}=F_H^{(0)}\vec{e}_x$ and $\vec{L}_H^{(0)}=L_H^{(0)}\vec{e}_y$, contain contributions due to the translational (T) and the rotational (R) motion, $F_H^{T,R}$ and $L_H^{T,R}$, respectively. They can be expressed as
\begin{subequations}
\begin{align}
F_H^{(0)} &= F_H^T+F_H^R = 6 \pi\mu a \left(U_\parallel \bar{F}^T + a\Omega^{(0)} \bar{F}^R \right), \label{eq:fH_0}\\
L_H^{(0)} &= L_H^T+L_H^R = 8 \pi\mu a^2 \left(U_\parallel \bar{L}^T + a\Omega^{(0)} \bar{L}^R \right),
\end{align}
\end{subequations}
where $\bar{F}^{T,R}$ and $\bar{L}^{T,R}$ denote the dimensionless hydrodynamic forces and torques.
Since the motion of the sphere is torque-free, $L^{(0)}_H=0$, the rotational velocity evaluates to
\begin{align}
\Omega^{(0)} = -\frac{U_\parallel}{a}\frac{\bar{L}^T}{\bar{L}^R}. \label{eq:omega0}
\end{align}
Similarly, the translational velocity, $U_\parallel$, is obtained from the force balance, $F_H^{(0)}+F=0$. Using the relation for the rotational velocity, Eq.~\eqref{eq:omega0}, we find that the hydrodynamic force reduces to
\begin{align}
F_H^{(0)} &=6 \pi\mu a U_\parallel \frac{\bar{F}^T \bar{L}^R-\bar{F}^R\bar{L}^T}{\bar{L}^R} \equiv 6 \pi\mu a U_\parallel \mathcal{R}_\parallel, \label{eq:force_parallel}
\end{align}
where we have introduced the dimensionless resistance along the direction of the force, $\mathcal{R}_\parallel$.

\section{Bispherical solution for the roughness-induced velocities \label{app:full}}
We summarize here the general solution for particle motion parallel (along the $x-$direction), transverse (along the $y-$direction), and perpendicular (along the $z-$direction) to a plane wall in terms of bispherical coordinates~\cite{ONeill:1964,Brenner:1961, Dean:1963}.
Therefore, we introduce the coordinates $(\eta,\xi)$ by
\begin{align}
r=c\frac{\sin\eta}{\cosh\xi-\cos\eta} \qquad \text{ and } \qquad z=c\frac{\sinh\xi}{\cosh\xi-\cos\eta},
\end{align}
where the parameter $c=a\sinh\alpha$ is determined by the boundary conditions and
\begin{align}
\alpha = \arccosh\left(\frac{h+a}{a}\right)=\ln\left(\frac{h+a}{a}+\sqrt{\left(\frac{h+a}{a}\right)^2-1}\right).
\end{align}
For the calculation of the fluid flow, it is convenient to introduce the rescaled variables
\begin{align}
r = c R, \quad z = cZ, \quad \vec{u}^{(0)} = U_\parallel \vec{U}^{(0)}, \quad\hat{\boldsymbol\sigma}^z= \frac{\mu \hat{U}_z a^2}{c^3}\hat{\boldsymbol\Sigma}^{z}, \quad \hat{\boldsymbol\sigma}^y = \frac{\mu \hat{U}_y}{c}\hat{\boldsymbol\Sigma}^{y}, \quad \hat{\vec{\sigma}}^{Ri}= \mu \hat{\Omega}_i \hat{\vec{\Sigma}}^{Ri} \text{ for } i=x,y,z.
\end{align}
In the following sections, we provide quantities, such as hydrodynamic forces and torques, and components of the velocity fields and the corresponding stress tensors, that are required for evaluating roughness-induced translational and rotational velocities, Eqs.~\eqref{eq:U_parallel}-\eqref{eq:U_perp} and Eqs.~\eqref{eq:omega_x}-\eqref{eq:omega_z}, respectively.

\subsection{Translational velocities}
\begin{enumerate}
\item \emph{Hydrodynamic forces.}
The hydrodynamic forces on a sphere moving parallel to the surface can be written as $\hat{F}_H^{x,y}=6\pi\mu a \mathcal{R}_\parallel \hat{U}_{x,y}$ with the dimensionless resistance (introduced in Eq.~\eqref{eq:force_parallel})
\begin{align}
\mathcal{R}_\parallel&=\frac{\bar{F}^T \bar{L}^R-\bar{F}^R\bar{L}^T}{\bar{L}^R},
\end{align}
that depends on forces and torques due to translation,
\begin{subequations}
\begin{align}
\bar{F}^{T}&=\frac{\sqrt{2}}{6}\sinh\alpha\sum_{n=0}^\infty E_n+n(n+1)C_n, \label{eq:f_T}\\
\begin{split}
\bar{L}^T &= \frac{1}{12\sqrt{2}}\sinh^2\alpha\sum_{n=0}^\infty (2+e^{-(2n+1)\alpha})(n(n+1)(2 A_n+C_n\coth\alpha)-(2n+1-\coth\alpha)E_n)\\
&\qquad\qquad\qquad\qquad\qquad +(2-e^{-(2n+1)\alpha})(n(n+1)B_n\coth\alpha -(2n+1-\coth\alpha)D_n)
, \label{eq:l_T}
\end{split}
\end{align}
\end{subequations}
and due to rotation,
\begin{subequations}
\begin{align}
\bar{F}^R &= -\frac{\sqrt{2}}{6}\sinh\alpha^2\sum_{n=0}^\infty E_n^\star+n(n+1)C_n^\star, \label{eq:f_R}\\
\begin{split}
\bar{L}^R &= \frac{1}{3}\bigl(1-\frac{1}{4\sqrt{2}}\sinh^3\alpha\sum_{n=0}^\infty (2+e^{-(2n+1)\alpha})(n(n+1)(2 A_n^\star+C_n^\star\coth\alpha)-(2n+1-\coth\alpha)E_n^\star)\\
&\qquad\qquad\qquad\qquad\qquad +(2-e^{-(2n+1)\alpha})(n(n+1)B_n^\star\coth\alpha -(2n+1-\coth\alpha)D_n^\star)\bigr)
, \label{eq:l_R}
\end{split}
\end{align}
\end{subequations}
The recurrence relations for the coefficients $A_n, B_n, C_n, D_n, E_n$, are provided in Ref.~\cite{ONeill:1964}, and the coefficients $A_n^\star, B_n^\star, C_n^\star, D_n^\star, E_n^\star$, are obtained from recurrence relations given in Ref.~\cite{Dean:1963}.

Similarly, the hydrodynamic force on a sphere moving perpendicular to the surface~\cite{ONeill:1964,Brenner:1961} is $\hat{F}_H^z = 6\pi\mu a \mathcal{R}_\parallel \hat{U}_z$ with dimensionless resistance,
\begin{align}
\mathcal{R}_{\perp}&= \frac{4}{3}\sinh\alpha\sum_{n=1}^\infty \frac{n(n+1)}{(2n-1)(2n+3)}\left[\frac{2\sinh((2n+1)\alpha)+(2n+1)\sinh2\alpha}{4\sinh^2((n+1/2)\alpha)-(2n+1)^2\sinh^2\alpha}-1\right].
\end{align}

\item \emph{Velocity field and stress tensors}
\begin{enumerate}
\item \emph{Translation along the $x-$axis} induces rotation around the $y$-axis.
The components for the velocity field induced by the motion of the sphere along the applied force are
\begin{subequations}
\begin{align}
U^{(0)}_{R} &= \frac{1}{2}\left(RQ_1+U_2+U_0-\frac{c}{a}\frac{\bar{L}^T}{\bar{L}^R}\left(RQ_1^\star+U_2^\star+U_0^\star\right)\right)\cos\vartheta \label{eq:vel_ZR}\\
U^{(0)}_{\vartheta}  &= \frac{1}{2}\left(U_2-U_0-\frac{c}{a}\frac{\bar{L}^T}{\bar{L}^R}\left(U_2^\star-U_0^\star\right)\right)\sin\vartheta.\label{eq:vel_Ztheta}
\end{align}
\end{subequations}
The contribution due to translation is encoded in the functions $Q_1$, $U_0$, $U_2$, which depend on $\eta$ and $\xi$ (see Eqs.~(7)-(9) in Ref.~\cite{ONeill:1964}), and similarly $Q_1^\star$, $U_0^\star$, $U_2^\star$ correspond to rotation (see Eqs.~(6)-(7) in Ref.~\cite{Dean:1963}).

To evaluate the integrals, Eqs.~\eqref{eq:U_parallel}-\eqref{eq:omega_z}, we require the gradient of the velocity field [Eq.~\eqref{eq:grad_vel0}] as input.  Therefore, we need to calculate the derivatives of the velocity components, Eqs.~\eqref{eq:vel_ZR}-\eqref{eq:vel_Ztheta}, which we denote by $U^{(0)}_{R;Z}$ and $U^{(0)}_{\vartheta;Z}$. 

\item \emph{Translation along the $y-$axis}. The components of the stress tensor relevant for evaluating Eq.~\eqref{eq:U_y} are
\begin{align}
\hat{\Sigma}^y_{ZR}\Bigr|_{Z=0} &= \frac{\partial U_R^y}{\partial Z}\Bigr|_{Z=0} \qquad \text{ and } \qquad
\hat{\Sigma}^y_{Z\vartheta}\Bigr|_{Z=0} =\frac{\partial U_\vartheta^y}{\partial Z}\Bigr|_{Z=0}
\end{align}
where the required components of the velocity field are slighlty modified from Eqs.~\eqref{eq:vel_ZR}-\eqref{eq:vel_Ztheta},
\begin{subequations}
\begin{align}
U_R^y &= \frac{1}{2}\left(RQ_1+U_2+U_0- \frac{c}{a}\frac{\bar{L}^T}{\bar{L}^R}\left(RQ_1^\star+U_2^\star+U_0^\star\right)\right)\sin\vartheta,\label{eq:vel_ZR_y}\\
U_\vartheta^y &= -\frac{1}{2}\left(U_2-U_0- \frac{c}{a}\frac{\bar{L}^T}{\bar{L}^R}\left(U_2^\star-U_0^\star\right)\right)\cos\vartheta,  \label{eq:vel_Ztheta_y}
\end{align}
\end{subequations}
with the same functions, $Q_1, U_0, U_2$ and $Q_1^\star, U_0^\star, U_2^\star$, as introduced above. Here, we have taken into account that translation along the positive $y-$axis induces rotation in the clockwise direction.

\item \emph{Translation along the $z-$axis} is axisymmetric and thus the sphere translates without rotating. The component of the stress tensor relevant for evaluating Eq.~\eqref{eq:U_perp} is obtained by
\begin{align}
\hat{\Sigma}_{ZR}^z\Bigr|_{Z=0} = \frac{1}{R}\frac{\partial^2 \Psi}{\partial Z^2}\Bigr|_{Z=0},
\end{align}
where we have introduced the rescaled stream function ($\psi = \hat{U}_z a^2 \Psi$) in terms of bispherical coordinates,
\begin{align}
\Psi(\eta,\xi) &= -(\cosh\xi-\mu)^{-3/2}\sum_{n=0}^\infty U_n(\xi)C_{n+1}^{-1/2}(\mu).
\end{align}
Here, we have abbreviated $\mu = \cos\eta$ and introduced
\begin{align}
U_n(\xi) = a_n\cosh((n-1/2)\xi)+b_n\sinh((n-1/2)\xi)+c_n\cosh((n+3/2)\xi)+d_n\sinh((n+3/2)\xi),
\end{align}
with coefficients $a_n, b_n, c_n, d_n$ dependent on $\alpha$ (see Eqs.~(2.16), (2.17) in Ref.~\cite{Brenner:1961}). Further, $C_{n+1}^{-1/2}(\mu)$ denotes the Gegenbauer polynomial of order $n+1$ and degree $-1/2$, which is related to the Legendre polynomials via the relation
$C_{n+1}^{-1/2}(\mu) = (P_{n-2}(\mu)-P_n(\mu))/(2n-1)$.
\end{enumerate}
\item \emph{Roughness-induced velocities.} Rescaling and inserting the expressions derived for the velocity gradient and the stresses into Eqs.\eqref{eq:U_parallel}-\eqref{eq:U_perp},  provides the first-order translational velocities
\begin{subequations}
\begin{align}
U_x^{(1)} &= -\frac{U_\parallel}{6\pi \mathcal{R}_\parallel}\int_0^\infty\!\!\!\int_0^{2\pi} H(R,\vartheta; \vec{r}_S)  \left[\hat{\Sigma}^x_{ZR} U^{(0)}_{R;Z}+\hat{\Sigma}^x_{Z\vartheta} U^{(0)}_{\vartheta;Z}\right]_{Z=0} \ R \ \diff\vartheta\diff R ,\label{eq:full_vel_final_para}\\
U_y^{(1)} &= -\frac{U_\parallel}{6\pi \mathcal{R}_\parallel}\int_0^\infty\!\!\!\int_0^{2\pi} H(R,\vartheta; \vec{r}_S)  \left[\hat{\Sigma}^y_{ZR} U^{(0)}_{R;Z}+\hat{\Sigma}^y_{Z\vartheta} U^{(0)}_{\vartheta;Z}\right]_{Z=0} \ R \ \diff\vartheta\diff R ,\label{eq:full_vel_final_y}\\
U_z^{(1)} &= -\frac{a^2U_\parallel }{6\pi c^2 \mathcal{R}_\perp}\int_0^\infty\!\!\!\int_0^{2\pi} H(R,\vartheta; \vec{r}_S)  \left[\hat{\Sigma}^z_{ZR} U^{(0)}_{R;Z}\right]_{Z=0} \ R \ \diff\vartheta\diff R. \label{eq:full_vel_final_perp}
\end{align}
\end{subequations}

\item \emph{Arbitrary force-direction.} The roughnness-induced velocities for the case of a force, $\vec{F}= F\vec{e}$ that is directed parallel to $S_0$ but along an arbitrary direction $\vec{e}= \cos\varphi_0 \vec{e}_x+ \sin\varphi_0\vec{e}_y$ with angle $\varphi_0$, can be merely obtained as a linear combination of the flow fields introduced above. In particular, we only require as input the modified zeroth-order flow-field, $\vec{u}^{(0)}\cos\varphi_0+\vec{u}^y\sin\varphi_0$ with $\vec{u}^{(0)}$ from
Eqs.~\eqref{eq:vel_ZR}-\eqref{eq:vel_Ztheta} and $\vec{u}^y$ from  Eqs.~\eqref{eq:vel_ZR_y}-\eqref{eq:vel_Ztheta_y}, and evaluate Eqs.~\eqref{eq:full_vel_final_para}-\eqref{eq:full_vel_final_perp} accordingly.
\end{enumerate}

\subsection{Rotational velocities}
The solutions for the velocity fields of the auxiliary problems can be to a large extent inferred from the equivalent translational problem. The velocities here are rescaled by $\vec{u}^{Rx,Ry}=c\hat{\Omega}\vec{U}^{Rx,Ry}$.
\begin{enumerate}
\item \emph{Hydrodynamic torques.} The torque for a sphere rotating around the $z-$axis can be expressed in terms of a dimensionless resistance,
$\hat{L}_H^z = 8\pi\mu a^3\hat{\Omega}\mathcal{R}^R_\perp$,
\begin{align}
\mathcal{R}^R_\perp &= \sum_{n=0}^\infty \frac{\sinh^3(\alpha)}{\sinh^3((m+1)\xi)}.
\end{align}
Similarly, a sphere that rotates with $\hat{\Omega}$ around an axis parallel to the wall also translates (if it is force-free). We calculate the translational velocity induced by rotation by transferring Eq.~\eqref{eq:fH_0} to the given problem ($\hat{F}_H=0$),
\begin{align}
U_\parallel &= -a\hat{\Omega}\frac{\bar{F}^R}{\bar{F}^T},
\end{align}
and we obtain the dimensionless resistance
\begin{align}
\mathcal{R}_\parallel^R &= \frac{\bar{L}^R\bar{F}^T-\bar{L}^T\bar{F}^R}{\bar{F}^T}.
\end{align}
The expressions for $\bar{F}^T,\bar{F}^R,\bar{L}^T, \bar{L}^R$ are provided in Eqs.~\eqref{eq:f_T}-\eqref{eq:l_R}.
\item \emph{Velocity field and stress tensors.}
\begin{enumerate}

\item \emph{Rotation around the $x-$axis} entails translation along the negative $y-$direction. Thus, the corresponding relevant (rescaled) components for the velocity field read
\begin{subequations}
\begin{align}
U_R^{Rx} &= -\frac{1}{2}\left(RQ_1^\star+U_2^\star+U_0^\star-\frac{a}{c}\frac{\bar{F}^R}{\bar{F}^T}\left(RQ_1+U_2+U_0\right)\right)\sin\vartheta,\\
U_\vartheta^{Rx} &= \frac{1}{2}\left(U_2^\star-U_0^\star-\frac{a}{c}\frac{\bar{F}^R}{\bar{F}^T}\left(U_2-U_0\right)\right)\cos\vartheta,
\end{align}
\end{subequations}
where we have used the same notation as above.

\item \emph{Rotation around the $y-$axis} induces translation along the $x-$direction and the (rescaled) velocity components read
\begin{subequations}
\begin{align}
U^{Ry}_{R} &= \frac{1}{2}\left(RQ_1^\star+U_2^\star+U_0^\star-\frac{a}{c}\frac{\bar{F}^R}{\bar{F}^T}\left(RQ_1+U_2+U_0\right)\right)\cos\vartheta,\\
U^{Ry}_{\vartheta}  &= \frac{1}{2}\left(U_2^\star-U_0^\star-\frac{a}{c}\frac{\bar{F}^R}{\bar{F}^T}\left(U_2-U_0\right)\right)\sin\vartheta,
\end{align}
\end{subequations}
where we have used the same notation as above.
\item \emph{Rotation around the $z-$axis} is axisymmetric and the relevant stress coefficient evaluates to~\cite{Jeffery:1915}
\begin{align}
\hat{\Sigma}^{Rz}_{Z\vartheta}&= \frac{\partial}{\partial Z} \left[\sqrt{\cosh \xi-\cos\eta}\sum_{n=1}^\infty B_n^\star\sinh((n+1/2)\xi) P_n^1(\cos\eta)\right]
\end{align}
with coefficients $B_n^\star$ that depend on $\xi$ (see Ref.~\cite{Jeffery:1915}).
\end{enumerate}
\item The \emph{roughness-induced velocities} for arbitrary surface shapes can be calculated by evaluating the integrals
\begin{subequations}
\begin{align}
\Omega_x^{(1)} &= -\frac{U_\parallel c}{8\pi a^2 \mathcal{R}^R_\parallel}\int_0^\infty\!\!\!\int_0^{2\pi} H(R,\vartheta; \vec{r}_S)  \left[\hat{\Sigma}^{Rx}_{ZR} U^{(0)}_{R;Z}+\hat{\Sigma}^{Rx}_{Z\vartheta} U^{(0)}_{\vartheta;Z}\right]_{Z=0} R \ \diff\vartheta \diff R, \label{eq:full_omega_x}\\
\Omega_y^{(1)} &= -\frac{U_\parallel c}{8\pi a^2 \mathcal{R}^R_\parallel}\int_0^\infty\!\!\!\int_0^{2\pi} H(R,\vartheta; \vec{r}_S)  \left[\hat{\Sigma}^{Ry}_{ZR} U^{(0)}_{R;Z}+\hat{\Sigma}^{Ry}_{Z\vartheta} U^{(0)}_{\vartheta;Z}\right]_{Z=0} R \ \diff\vartheta \diff R, \label{eq:full_omega_y}\\
\Omega_z^{(1)} &= -\frac{U_\parallel c}{8\pi a^2 \mathcal{R}^R_\perp}\int_0^\infty\!\!\!\int_0^{2\pi} H(R,\vartheta; \vec{r}_S)  \left[\hat{\Sigma}^{Rz}_{Z\vartheta} U^{(0)}_{\vartheta;Z}\right]_{Z=0} R \ \diff\vartheta \diff R. \label{eq:full_omega_z}
\end{align}
\end{subequations}
\end{enumerate}
As discussed above, also the rotational velocities for a sphere driven along an arbitrary direction are obtained by exchanging the zeroth-order velocity gradient.

\section{Random rough surface: variance}
In the local cylindrical coordinate system, the surface shape [Eq.~$(7)$ in main text] is written
\begin{align}
H(R,\vartheta;\mathbf{r}_S) =\frac{1}{N}\!\!\sum_{n,m=1}^N\!\!\!\alpha_{nm}\sin\left(\vec{k}^n_{m}\!\cdot\!\left(\vec{r}_S+cR\vec{e}_R\right)\right)+\beta_{nm} \cos\left(\vec{k}^n_{m}\!\cdot\!\left(\vec{r}_S+cR\vec{e}_R\right)\right),
\label{eq:H_rough_c}
\end{align}
with $\vec{e}_R = (\cos\vartheta,\sin\vartheta)^T$ and $\vec{r}_S=(x_S,y_S)^T$ the position of the sphere relative to the underlying surface.
The roughness-induced velocities are obtained by inserting Eq.~\eqref{eq:H_rough_c} into Eqs.~\eqref{eq:full_vel_final_para}-\eqref{eq:full_vel_final_perp}.

The general expression for the $x$-component of the variance reads:
\begin{align}
\left\langle\left(U_x^{(1)}\right)^2\right\rangle &= \left(\frac{U_\parallel}{6\pi \mathcal{R}_\parallel N}\right)^2\left\langle \left[\int H(R,\vartheta; \vec{r}_S)  \left[\hat{\Sigma}^x_{ZR} U^{(0)}_{R;Z}+\hat{\Sigma}^x_{Z\vartheta} U^{(0)}_{\vartheta;Z} \right]_{Z=0} \ \diff S \right]^2 \right\rangle, \label{eq:full_var}
\end{align}
where we have abbreviated the integrals by $\diff S = R\diff R\diff\vartheta$. Exploiting the statistical properties of the surface amplitudes, $\langle\alpha_{nm}\alpha_{ij}\rangle=\langle\beta_{nm}\beta_{ij}\rangle=\delta_{ni}\delta_{mj}$, and expanding the expression of the surface shape [Eq.~\eqref{eq:H_rough_c}],  the integral simplifies to
\begin{align}
\begin{split}
\left\langle\left(U_x^{(1)}\right)^2\right\rangle = \left(\frac{U_\parallel}{6\pi \mathcal{R}_\parallel N}\right)^2 \sum_{n,m}^N \Biggl[&\left(\int  \sin(cR\vec{k}_{n}^m\cdot\vec{e}_R) \left[\hat{\Sigma}^x_{ZR} U^{(0)}_{R;Z}+\hat{\Sigma}^x_{Z\vartheta} U^{(0)}_{\vartheta;Z}\right]_{Z=0} \ \diff S\right)^2\\
& +\left(\int  \cos(cR\vec{k}_{n}^m\cdot\vec{e}_R) \left[\hat{\Sigma}^x_{ZR} U^{(0)}_{R;Z}+\hat{\Sigma}^x_{Z\vartheta} U^{(0)}_{\vartheta;Z}\right]_{Z=0} \ \diff S\right)^2\Biggr]. \label{eq:full_var_s}
\end{split}
\end{align}
In particular, we find that cross-terms  and thus particle positions cancel out. Similarly, we find the variances, $\left\langle\left(U_y^{(1)}\right)^2\right\rangle$ and $\left\langle\left(U_z^{(1)}\right)^2\right\rangle$, and the variances for the rotational velocities.

\section{Validation of roughness-induced velocities with a boundary integral method}
We have validated our theoretical predictions with results obtained from a boundary integral method, which takes into account the surface shape explicitly.

\emph{Boundary integral method (BIM).}
We solve the single-layer boundary integral
equations~\cite{Pozrikidis:1992} by employing a boundary element method. Zeroth-order quadrilateral elements are used to discretize the particle surface and the rough wall;
a six-patch structured mesh is generated on the former by mapping a cubic to a spherical surface; the horizontal extension of the rough wall is $50$ particle radii in each direction. By dividing each element into four triangular sub-elements, singular integrations are performed based on the polar coordinates transformation~\cite{Pozrikidis:2002} and Gauss-Legendre quadrature. Adaptive mesh refinement is implemented to facilitate a more accurate computation of the sharp nearly-singular integrations emerging at a small particle-wall distance. The numerical setup has been developed and adapted to study the hydrodynamics of a
particle inside a tube~\cite{Zhu:2013} and inside a deformable fluid interface~\cite{Zhu:2017}.

\begin{figure*}[htp]
\centering
\includegraphics[width = \linewidth]{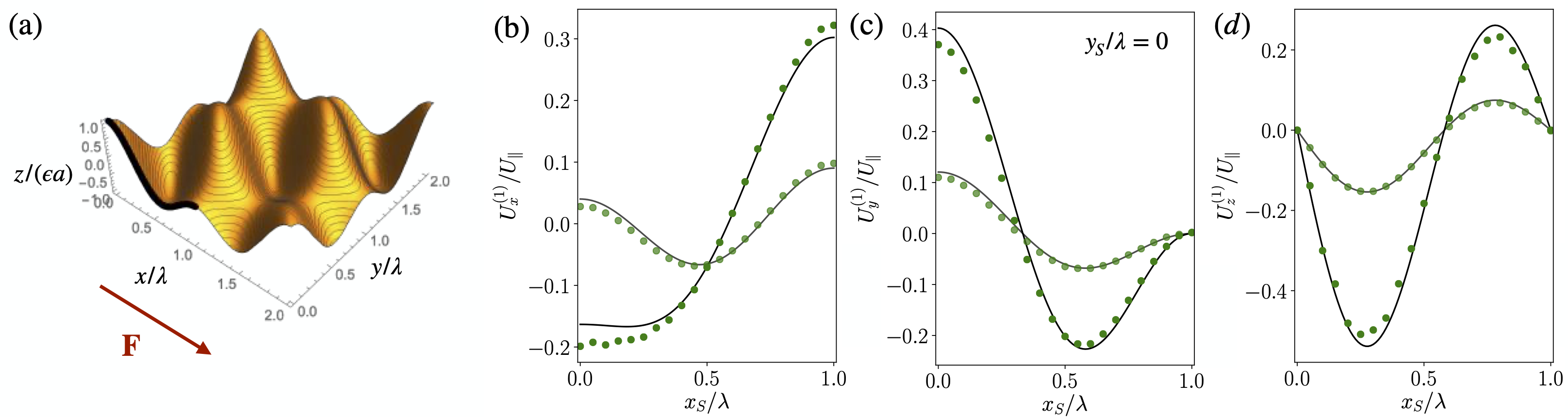}
\caption{\label{fig:validation_trans_bim} (a) Periodic surface shape  and roughness-induced velocities, $U^{(1)}_{x,y,z}$, along the (b) $x-$, (c) $y-$, and (d) $z-$direction.
The velocities are evaluated at different particle positions $x_S,y_S$, which are indicated as black line in (a), and for differnt particle-surface distances, $h/a$.
Lines indicate analytical predictions and symbols
are resuls from the boundary integral method, where the roughness parameter is $\epsilon = 0.02$ and the wavenumber is $\lambda/a = 2$.
}
\end{figure*}

\emph{Comparison between analytical solution and BIM.}
We calculated the translational and rotational roughness-induced velocities for a sphere near a periodic surface which is exposed to an external force, $\vec{F}=F\vec{e}_x$ (shown in Fig.~\ref{fig:validation_trans_bim}(a)) with shape
$H(x,y)= \left(\cos\left(\frac{k}{2}\left(x+2y\right)\right)+\cos\left(\frac{k}{2}\left(2x+y\right)\right)\right)/2$.

\begin{figure*}[htp]
\centering
\includegraphics[width = 0.85\linewidth]{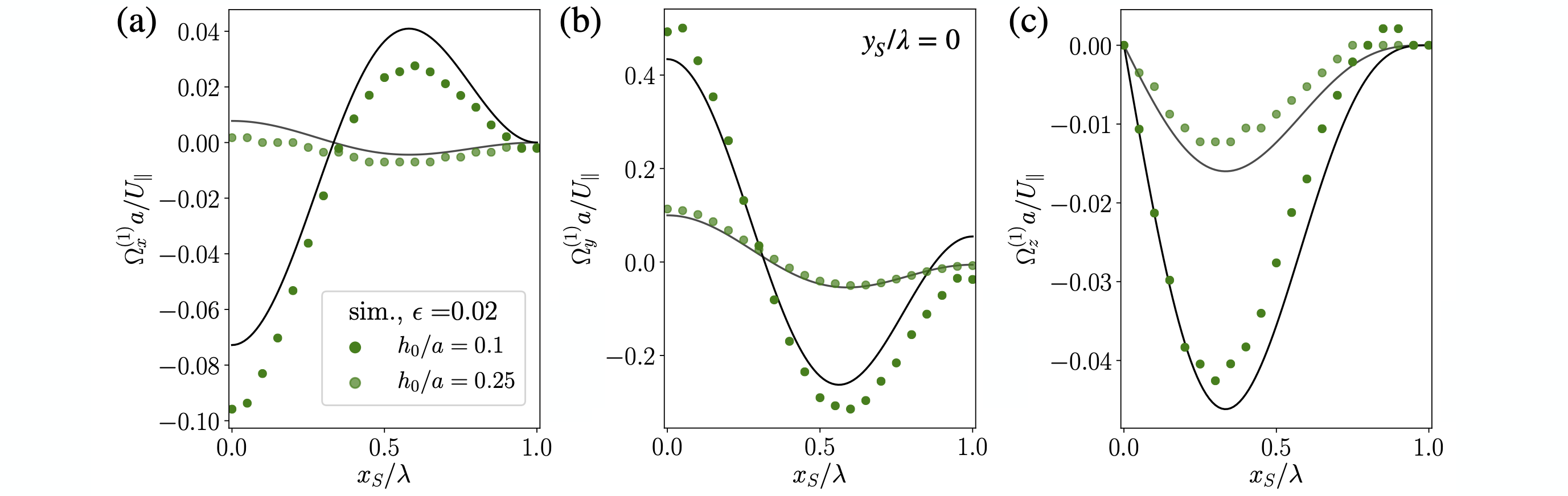}
\caption{\label{fig:validation_rot_bim} Roughness-induced rotational velocities, $\Omega^{(1)}_{x,y,z}$, along the (a) $x-$, (b) $y-$, and (c) $z-$axes.
The velocities are evaluated at different particle positions $x_S,y_S$, which are indicated as black line in Fig.~\ref{fig:validation_trans_bim}~(a), and for differnt particle-surface distances, $h/a$.
Lines indicate analytical predictions and symbols
are resuls from the boundary integral method, where the roughness parameter is $\epsilon = 0.02$ and the wavenumber is $\lambda/a = 2$.
}
\end{figure*}

The translational and rotational velocities are shown in Figs.~\ref{fig:validation_trans_bim} and \ref{fig:validation_rot_bim}, respectively. In particular, the analytical predictions for the translational velocities show good agreement with the numerical solutions. The rotational velocities, in particular, along the $x-$ and $z-$ direction show some deviations from the numerical results which could be amongst other factors due to the accuracy of the method, as these are one order in magnitude smaller than the others and thus more difficult to resolve.


\bibliography{final_files_arxiv/final_manuscript/literature}